\documentclass[preprint]{revtex4}
\usepackage{graphicx}
\usepackage{bm}
\usepackage{dcolumn}
\usepackage{amsmath}
\usepackage{amsmath}
\usepackage{amssymb}
\usepackage{color}

\usepackage{epsfig}
\begin{document}
\newcommand{\newc}{\newcommand}
\newc{\ra}{\rightarrow}
\newc{\lra}{\leftrightarrow}
\newc{\lsim}{\buildrel{<}\over{\sim}}
\newc{\gsim}{\buildrel{>}\over{\sim}}
\title{Constraints on the $\Lambda$CDM model with redshift tomography}
\author{Rong-Gen Cai$^{1}$}
\email{cairg@itp.ac.cn}
\author{Zong-Kuan Guo$^{1}$}
\email{guozk@itp.ac.cn}
\author{Bo Tang$^{1}$}
\email{tangbo@itp.ac.cn}

\affiliation{
    $^{1}$ State Key Laboratory of Theoretical Physics, Institute of
Theoretical Physics, Chinese Academy of Sciences, P.O. Box 2735,
Beijing 100190, China.   \\
}
\date{\today}

\begin{abstract}
Recently released Planck data favor a lower value of the Hubble
constant and a higher value of the fraction matter density in the
standard $\Lambda$CDM model, which are discrepant with some of the
low-redshift measurements. Within the context of this cosmology, we
examine the consistency of the estimated values for the Hubble
constant and fraction matter density with redshift tomography. Using
the SNe Ia, Hubble parameter, BAO and reduced CMB data, which are divided
into three bins, we find no statistical evidence for any tension in
the three redshift bins.
\end{abstract}

\maketitle
\section{Introduction}

More than one decade ago it was found that our universe is in an
accelerating expansion based on the distance measurement of type Ia
supernovae (SNe Ia)~\cite{Riess:1998cb,Perlmutter:1998np}. This
observation is consistent with other astronomical observations such
as Hubble parameter, large scale structure and cosmic microwave
background radiation (CMB), etc. To explain this accelerating
expansion, one has to introduce the so-called dark energy with
negative pressure in the general relativity framework, or to modify
the general relativity at cosmic scales. Although suffered from some
theoretical issues, the cosmological
constant~\cite{Sahni:2002kh,Padmanabhan:2002ji} introduced by
Einstein himself in 1917 is the most simple and economical candidate
for the dark energy. Indeed the standard $\Lambda$CDM model turns
out to be consistent with several precise astronomical observations,
such as SNe Ia~\cite{Suzuki:2011hu}, Wilkinson Microwave Anisotropy
Probe (WMAP) measurements of CMB~\cite{Hinshaw:2012aka}, and baryon
acoustic oscillation, etc. If the standard $\Lambda$CDM model
properly describes our universe,  the current Hubble constant $H_0$
and fraction matter density $\Omega_{m0}$ should be consistent with
those estimated by different observations made at different
redshifts.

However, the recently released Planck data~\cite{Ade:2013zuv} favor
a higher value of $\Omega_{m0}=0.315\pm 0.017$ and a lower value of
$H_0=(67.3\pm 1.2)$ km s$^{-1}$ Mpc$^{-1}$ in the standard
six-parameter $\Lambda$CDM cosmology, obtained by using Planck+WP,
where WP stands for WMAP polarization data. These values are in
tension with the magnitude-redshift relation for SNe Ia and recent
direct measurements of $H_0$, such as the Hubble space telescope
observations of Cepheid variables with $H_0=(73.8\pm 2.4)$ km
s$^{-1}$ Mpc$^{-1}$~\cite{Riess:2011yx} and $H_0=[74.3\pm 1.5({\rm
stat.})\pm 2.1({\rm sys.})]$ km s$^{-1}$ Mpc$^{-1}$ obtained by
using a mid-infrared calibration of the Cepheid distance scale based
on observations at 3.6 $\mu m$ with the Spitzer Space
Telescope~\cite{Freedman:2012ny}. Of course, if relax the
restriction of the standard six-parameter $\Lambda$CDM model, for
example, consider the dynamical dark energy model~\cite{Xia:2013dea}
or include the dark radiation~\cite{Cheng:2013csa}, the tension
might be alleviated. In \cite{Hu:2013aqa},  Hu  {\it et al.} found that there is another
way to alleviate this tension in modified gravity models. Furthermore it was reported in \cite{Li:2013dwy} that
this tension may also be alleviated by if one first calibrates the
light-curve fitting parameters in the distance estimation in SNe Ia observations with
the angular diameter distance data of the galaxy clusters, with the help of the distance-duality relation.
Very recently,
Efstathiou~\cite{Efstathiou:2013via} reanalyzed the Cepheid data and
found $H_0=(70.6\pm 3.3)$ km s$^{-1}$ Mpc$^{-1}$ based on the NGC
4258 maser distance and $H_0=(72.5\pm 2.5)$ km s$^{-1}$ Mpc$^{-1}$
with three distance anchors combined, which alleviates the tension
compared to the result obtained by Riess {\it et
al.}~\cite{Riess:2011yx}, but the latter still differs by 1.9$\sigma$
from the Planck value. In addition, by comparing the eight
ultra low redshift SNe Ia data $(z=0.0043~{\rm
to}~0.0072)$~\cite{Riess:2011yx}, with low redshift data $(z<0.04)$ from
the Union2.1 compilation~\cite{Union2.1} and Planck
data~\cite{Ade:2013zuv}, Zhang and Ma found that the present
expansion of the universe estimated from the low redshift
measurements is higher than the one estimated from high redshift
observations in the $\Lambda$CDM model~\cite{Zhang:2013hma}. In
other words, higher redshift measurements give a lower value of
$h$, the reduced Hubble constant.

These discrepancies seemingly imply that the standard $\Lambda$CDM
model cannot well describe the properties of the universe at all
redshift if the major sources of systematic errors of these
observations have been controlled.
 In
this paper we detect these discrepancies in the $\Lambda$CDM model
with redshift tomography. We divide the redshift range under
consideration into three bins and use observation data in each bin
to separately constrain the Hubble constant and fraction matter
density in the $\Lambda$CDM model. In the literature the redshift
tomography is often used to see the dynamical property of dark
energy by piecewise parametrization of the equation of state of the
dark energy. Here our goal is to see the consistency of the
$\Lambda$CDM model at different redshifts, therefore we focus on the
$\Lambda$CDM model. The data sets we use here include the Union$2.1$
SNe Ia data~\cite{Union2.1}, 19
Hubble parameter  $H(z)$ data~\cite{Simon:2004tf,Stern:2009ep,Moresco:2012jh},
Baryon Acoustic Oscillation (BAO) data
measured by the 6 degree Field Galaxy Survey (6dFGS), SDSS DR7, SDSS
DR9 and WiggleZ surveys,  reduced nine-year WMAP data
(WMAP9) and reduced Planck data both based on the flat $\Lambda$CDM model.

The paper is organized as follows. In section~\ref{sec:model} we describe the redshift tomography method and observational data.
In section~\ref{sec:results} we show the results of different combination of data sets to constrain the base $\Lambda$CDM model based on the SNe Ia data and the redshift tomography analysis.
The results are summarized in section~\ref{sec:conclusions}.


\section{Method and data \label{sec:model}}

In a spatially flat Friedmann-Robertson-Walker universe, the Hubble
parameter is given by the Friedmann equation
\begin{equation}
H^2(z)=H^2_0\left[\Omega_{r0}(1+z)^4+\Omega_{dm0}(1+z)^3+\Omega_{b0}(1+z)^3+(1-\Omega_{m0}-\Omega_{r_0})\right],
\end{equation}
for the $\Lambda$CDM model, where the redshift $z$ is defined by
$(1+z)=1/a$, and $\Omega_{r0}$, $\Omega_{dm0}$ and $\Omega_{b0}$ are
the present values of the fraction energy density  for radiation,
dark matter and baryon matter, respectively. The latter two are
often written as the total matter density
$\Omega_{m0}=\Omega_{b0}+\Omega_{dm0}$. The radiation density is the
sum of photons and relativistic neutrinos~\cite{Hinshaw:2012aka}:
\begin{equation}
\Omega_{r0}=\Omega^{(0)}_{\gamma}(1+0.2271N_{eff}),
\end{equation}
where $N_{eff}=3.046$ is the effective number of neutrino species  in the Standard Model~\cite{Mangano:2005cc}, and $\Omega^{(0)}_{\gamma}=2.469\times10^{-5}h^{-2}$ for $T_{\rm CMB}=2.725K$
($h \equiv H_0/100$ km s$^{-1}$ Mpc$^{-1}$).

We focus on constraints on the Hubble constant and the fraction
matter density  in the context of the $\Lambda$CDM cosmology from
the low-redshift observational data including Union$2.1$ SNe Ia
sample, Hubble parameter and BAO data, in combination with the
high-redshift CMB measurements. We adopt a redshift tomography
method to examine the flat $\Lambda$CDM model. Since the SNe Ia data
cannot alone constrain the $\Lambda$CDM model very well,  it could
be even worse in each redshift bin because of the decreasing of data
points (so do the Hubble parameters), we then divide the redshift
into three bins so that the BAO data can distribute uniformly in the
first two bins, while the CMB data are in the third bin.  As a
result, these data are divided into three combinations in the
following redshift bins: $0-0.28$, $0.28-0.73$ and $>0.73$. The
distribution of data is listed in Table~\ref{tab:distribution}. To
see the difference, we will use WMAP9 and Planck data separately.¡¡

\begin{table}[!th]
\begin{tabular}{l|l|l|l|l}
\hline\hline redshift bin  & SNe Ia  & Hubble & BAO & CMB
\tabularnewline \hline $0-0.28$              & $283$ & $5$ & 6dF,
DR7a, RBAO1 & -- \tabularnewline \hline $0.28-0.73$           &
$212$ & $5$ & DR7-re, WiggleZ, DR9, RBAO2 & -- \tabularnewline
\hline $>0.73$ & $85$ & $9$ & -- & WMAP9/Planck \tabularnewline
\hline
\end{tabular}
\caption{\label{tab:distribution}Distribution of SNe Ia, Hubble,
BAO, CMB data in three redshift bins.}
\end{table}

The best-fit values of $\Omega_{m0}$ and $h$, and their 68\% and
95\% confidence level (CL) errors are obtained by performing the
Markov Chain Monte Carlo analysis in the multidimensional parameter
space in a Bayesian framework. Since the Hubble constant is
completely degenerate with the absolute magnitude of SNe Ia, SNe Ia
data are not sensitive to the Hubble constant. Therefore, in our
analysis we marginalize analytically over the Hubble constant when
the SNe Ia data are concerned. Moreover, note that the fraction
baryon energy density $\Omega_{b0}$ is involved in the likelihood
for the BAO and CMB data.


\subsection{Type Ia Supernovae}

The SNe Ia data set is an important tool to understand the evolution
of the universe. In this work, we adopt the Union$2.1$
compilation~\cite{Union2.1}, containing $580$ SNe Ia data over the
redshift range $0.015 \le z \le 1.414$. The chisquare is defined as
as
\begin{equation}
\chi^2_{SN}=\sum_{i=1}^{N}\frac{[\mu^{obs}(z_i)-\mu^{th}(z_i)]^2}{\sigma_{SN}^2(z_i)} ,
\end{equation}
where $N$ is the data number in the redshift interval we are interested in, $\mu^{obs}(z)$ is the measured distance modulus from the data and $\mu^{th}(z)$ is the theoretical distance modulus, defined as
\begin{equation}
\mu^{th}(z)=5\log_{10}{d_L}+\mu_0,~~\mu_0=42.384-5\log_{10}{h}.
\end{equation}
The luminosity distance is
\begin{equation}
d_L(z)=(1+z)\int^x_0\frac{dx}{E(x)},
\end{equation}
where $E(z) \equiv H(z)/H_0$. We can eliminate the nuisance
parameter $\mu_0$ by expanding $\chi^2$ with respect to
 $\mu_0$~\cite{Nesseris:2005ur} :
\begin{equation}
\chi^2_{SN}=A+2B\mu_0+C\mu^2_0 ,
\end{equation}
where
\begin{equation}
\begin{split}
&A=\sum_{i=1}^{N}\frac{[\mu^{th}(z_i;\mu_0=0)-\mu^{obs}(z_i)]^2}{\sigma_{SN}^2(z_i)},\\
&B=\sum_{i=1}^{N}\frac{\mu^{th}(z_i;\mu_0=0)-\mu^{obs}(z_i)}{\sigma_{SN}^2(z_i)},\\
&C=\sum_{i=1}^{N}\frac{1}{\sigma_{SN}^2(z_i)} .
\end{split}
\end{equation}
The \textbf{$\chi^{2}_{SN}$} has a minimum as
\begin{equation}
\tilde{\chi}^{2}_{SN}=A-B^2/C~,\label{sn}
 \end{equation}
which is
independent of $\mu_{0}$. This technique is equivalent to performing a
uniform marginalization over $\mu_{0}$~\cite{Nesseris:2005ur}. We will adopt
\textbf{$\tilde{\chi}^{2}_{SN}$} as the goodness of fitting instead of \textbf{$\chi^{2}_{SN}$}.


\subsection{Observational Hubble parameter (HUB)}

The observational Hubble parameter can be obtained by using the
differential ages of passively evolving galaxies as
\begin{equation}
\begin{split}
H=-\frac{1}{1+z} \frac{\Delta z}{\Delta t}.
\end{split}
\end{equation}
We use 19 observational Hubble data over the redshift range $0.07\leq z \leq 2.3$,
which contain 11 observational Hubble data obtained from the differential ages of
passively evolving galaxies~\cite{Simon:2004tf,Stern:2009ep}, and 8 $H(z)$ data at eight different redshifts
obtained from the differential spectroscropic evolution of early type galaxies as a function of redshift~\cite{Moresco:2012jh}.
The chisqure is defined as
\begin{equation}
\chi^{2}_{HUB}=\sum_{i=1}^{N}\frac{[H_{th}(z_i)-H_{obs}(z_i)]^2}{\sigma_H^2(z_i)},
\end{equation}
where $H_{th}(z)$ and $H_{obs}(z)$ are the theoretical and observed values of Hubble parameter,
and $\sigma_H$ denotes the error bar of observed data.


\subsection{Baryon Acoustic Oscillation}

As a ruler to measure the distance-redshift relation, Baryon
Acoustic Oscillation provides an efficient method for measuring the
expansion history of the universe by using features in the cluster
of galaxies with large scale surveys. Here we use the results from
the following five BAO surveys: the 6dF Galaxy Survey, SDSS DR7,
SDSS DR9, WiggleZ measurements and the radial BAO measurement.

\subsubsection{6dF Galaxy Survey}

The 6dFGS BAO detection allows us to constrain the distance-redshift
relation at $z_{eff}=0.106$~\cite{Beutler:2011hx}. The low effective
redshift of 6dFGS makes it a competitive and independent alternative
to Cepheids and low redshift  supernovae in constraining the Hubble
constant. They achieved a measurement of the distance ratio
\begin{equation}
\frac{r_s(z_d)}{D_V(z=0.106)}=0.336\pm0.015,
\end{equation}
where $r_s(z_d)$ is the comoving sound horizon at the baryon drag epoch
when baryons became dynamically decoupled from photons.
The redshift $z_d$ is well approximated by~\cite{Eisenstein:1997ik}
\begin{equation}
z_d=\frac{1291(\Omega_{m0} h^2)^{0.251}}{1+0.659(\Omega_{m0} h^2)^{0.828}} [1+b_1(\Omega_{b0} h^2)^{b_2}],
\end{equation}
where
\begin{equation}
\begin{split}
&b_1=0.313(\Omega_{m0} h^2)^{-0.419}[1+0.607(\Omega_{m0} h^2)^{0.674}],\\
&b_2=0.238(\Omega_{m0} h^2)^{0.223}.
\end{split}
\end{equation}
The effective ``volume'' distance $D_V$ is a combination of the angular-diameter distance $D_A(z)$ and the Hubble parameter $H(z)$,
\begin{equation}
\begin{split}
D_V(z)&=\left [(\int^z_0\frac{dx}{H(x)})^2\frac{z}{H(z)}\right ]^{1/3}\\
&=[(1+z)^2D_A(z)^2\frac{z}{H(z)}]^{1/3}.
\end{split}
\end{equation}
The $\chi^2_{6dF}$ is given by
\begin{equation}
\chi^{2}_{6dF}=\frac{[({r_s(z_d)}/{D_V(0.106))_{th}}-0.336]^2}{0.015^2}.
\end{equation}

\subsubsection{SDSS DR7}

 The joint analysis of the 2-degree Field Galaxy Redshift
Survey data and the  Sloan Digital Sky Survey Data Release $7$ data
gives the distance ratio at $z=0.2$ and
$z=0.35$~\cite{Percival:2009xn}:
\begin{equation}
\begin{split}
\frac{r_s(z_d)}{D_V(z=0.2)}=0.1905\pm0.0061,\\
\frac{r_s(z_d)}{D_V(z=0.35)}=0.1097\pm0.0036.
\end{split}
\end{equation}
When the two data points are in the same redshift bin, we adopt the
$\chi^2_{DR7}$ given by
\begin{equation}
\chi^2_{DR7}=X^TV^{-1}X,
\end{equation}
where
\begin{equation}
X=
  \begin{bmatrix}
   [\frac{r_s(z_d)}{D_V(0.2)}]_{th}-0.1905\\
   [\frac{r_s(z_d)}{D_V(0.35)}]_{th}-0.1097
  \end{bmatrix},
\end{equation}
and the inverse covariance matrix is
\begin{equation}
V^{-1}=
  \begin{bmatrix}
   30124.1&-17226.9\\
   -17226.9&86976.6
  \end{bmatrix}.
\end{equation}
On the other hand, when the two data points are in the different
redshift bin, their $\chi^2_{DR7}$ are respectively given by
\begin{equation}
\begin{split}
&\chi^2_{DR7a}=\frac{[(\frac{r_s(z_d)}{D_V(0.2)})_{th}-0.1905]^2}{0.0061^2},\\
&\chi^2_{DR7b}=\frac{[(\frac{r_s(z_d)}{D_V(0.35)})_{th}-0.1097]^2}{0.0036^2}.
\end{split}
\end{equation}

\subsubsection{\normalsize{SDSS DR7 reanalysis}}

By applying the reconstruction technique~\cite{Eisenstein:2006nk} to the clustering of galaxies from
the SDSS DR7 Luminous Red Galaxies sample, and sharpening the BAO feature, Padmanabhan {\it et al.} obtained
 the distance ratio at $z=0.35$~\cite{Padmanabhan:2012hf} :
\begin{equation}
\frac{r_s(z_d)}{D_V(z=0.35)}=0.1126\pm0.0022.
\end{equation}
The $\chi^2_{DR7-re}$ used in the Markov Chain  Monte Carlo analysis is
\begin{equation}
\chi^2_{DR7re}=\frac{[(\frac{r_s(z_d)}{D_V(0.35)})_{th}-0.1126]^2}{0.0022^2}.
\end{equation}
Since the SDSS DR7 and SDSS DR7 reanalysis results are based on the same survey
and the latter gives a higher precision than the former, we include the SDSS DR7 reanalysis data when we
do the whole redshift analysis but not both together. On the other hand, when we do redshift tomography, we may refer
to part of the SDSS DR7 data at  $z=0.2$ and when the redshift bin contains $z=0.35$, we will use the SDSS DR7 reanalysis data.

\subsubsection{SDSS DR9}

The SDSS DR9 measurement at $z=0.57$ analyzed by Anderson {\it et
al.}~\cite{Anderson:2012sa} gives
\begin{equation}
\frac{r_s(z_d)}{D_V(z=0.57)}=0.0732\pm0.0012,
\end{equation}
which is the most precise determination of the acoustic oscillation
scale to date. The chisquare is defined as
\begin{equation}
\chi^2_{DR9}=\frac{[(\frac{r_s(z_d)}{D_V(0.57)})_{th}-0.0732]^2}{0.0012^2}.
\end{equation}

\subsubsection{The WiggleZ measurements}

The WiggleZ team encodes some shape information on the power
spectrum to measure the acoustic parameter~\cite{Blake:2011en}:
\begin{equation}
A(z)=\frac{D_V(z)\sqrt{\Omega_{m0}H_0}}{z}.
\end{equation}
The measurements of the baryon acoustic peak at redshifts $z=0.44$,
 $0.6$ and $0.73$ in the galaxy correlation function of the final
dataset of the WiggleZ Dark Energy Survey give the acoustic
parameter:
\begin{equation}
\begin{split}
&A(z=0.44)=0.474\pm0.034,\\
&A(z=0.60)=0.442\pm0.020,\\
&A(z=0.73)=0.424\pm0.021.
\end{split}
\end{equation}
The chisquare is defined as
\begin{equation}
\chi^2_{Wig}=X^TV^{-1}X,
\end{equation}
where
\begin{equation}
X=
  \begin{bmatrix}
  A(z=0.44)_{th}-0.474\\
  A(z=0.60)_{th}-0.442\\
  A(z=0.73)_{th}-0.424
  \end{bmatrix},
\end{equation}
and its inverse covariance matrix is
\begin{equation}
V^{-1}=
  \begin{bmatrix}
   1040.3&-807.5&336.8\\
   -807.5&3720.3&-1551.9\\
   336.8&-1551.9&2914.9
  \end{bmatrix}.
\end{equation}

\subsubsection{Radial BAO}

The radial (line-of-sight) baryon acoustic scale can also be
measured by using the SDSS data. It is independent from the BAO
measurements described above, which are averaged over all directions
or in the transverse directions. The measured quantity is
\begin{equation}
\bigtriangleup_z(z)=H(z)r_s(z_d),
\end{equation}
whose values are given by~\cite{Gaztanaga:2008de}
\begin{equation}
\begin{split}
&\bigtriangleup_z(0.24)=0.0407\pm{0.0011},\\
&\bigtriangleup_z(0.43)=0.0442\pm{0.0015}.
\end{split}
\end{equation}

\subsection{Cosmic Microwave Background}

In the CMB measurement, the distance to the last scattering surface can be accurately determined
from the locations of peaks and troughs of acoustic oscillations.
There are two quantities: one is the ``acoustic scale"
\begin{equation}
l_A=(1+z_*)\frac{\pi D_A(z_*)}{r_s(z_*)},
\end{equation}
and the other is the ``shift parameter"
 \begin{equation}
R=\sqrt{\Omega_{m0} H_0^2}(1+z_*)D_A(z_*).
\end{equation}
These quantities can be used to constrain cosmological parameters
without need to use the full data of WMAP9~\cite{Hinshaw:2012aka}.
Here $z_*$ is the redshift at the last scattering
surface~\cite{Hu:1995en}
\begin{equation}
z_*=1048[1+0.00124(\Omega_{b0} h^2)^{-0.738}][1+g_1(\Omega_{m0} h^2)^{g_2}],
\end{equation}
where
\begin{equation}
\begin{split}
&g_1=\frac{0.0783(\Omega_{b0} h^2)^{-0.238}}{1+39.5(\Omega_{b0} h^2)^{0.763}},\\
&g_2=\frac{0.560}{1+21.1(\Omega_{b0} h^2)^{1.81}}.
\end{split}
\end{equation}

Wang and Wang~\cite{Wang:2013mha} have obtained the mean values and covariance matrix of
${\{R, l_A, \Omega_{b0} h^2,n_s\}}$ from WMAP9 and Planck data respectively, based on the
 $\Lambda CDM$ model without assuming a flat universe. On the other hand, Shafer and
Huterer~\cite{Shafer:2013pxa} derived the related results about ${\{R, l_A, z_*\} }$ from WMAP9 and Planck data respectively, based on the flat $wCDM$ model. For our propose, following \cite{Wang:2013mha} and \cite{Shafer:2013pxa},
we first extract the mean values and covariance matrix of
${\{R, l_A, z_*\} }$ from WMAP9 and Planck data respectively based on a flat $\Lambda CDM$ model.

\subsubsection{WMAP9}

By using the WMAP9 data, we obtain the mean values for ${\{R, l_A, z_*\} }$ as
\begin{equation}
\langle l_A \rangle =301.95,~~~
\langle R \rangle =1.7257,~~~
\langle z_*\rangle =1088.96.
\end{equation}
Their inverse covariance matrix is
\begin{equation}
C_{WMAP9}^{-1}=
  \begin{bmatrix}
   3.087&15.160&-1.456\\
   15.160&12805.3&-217.021\\
   -1.456&-217.021&5.552
  \end{bmatrix}.
\end{equation}
The chisquare for the reduced WMAP9 data is defined by
\begin{equation}
\chi^2_{WMAP9}=X^TC_{WMAP9}^{-1}X,
\end{equation}
where
\begin{equation}
X=
  \begin{bmatrix}
  l_A-301.95\\
  R-1.7257\\
  z_*-1088.96.
  \end{bmatrix}
\end{equation}


\subsubsection{Planck}

By using the Planck data, we obtain the mean values for ${\{R, l_A, z_*\} }$ as
\begin{equation}
\langle l_A \rangle =301.65,~~~
\langle R \rangle =1.7500,~~~
\langle z_*\rangle =1090.33.
\end{equation}
Their inverse covariance matrix is
\begin{equation}
C_{Planck}^{-1}=
  \begin{bmatrix}
   40.909&-405.455&-0.5443\\
   -405.455&55662.8&-751.123\\
   -0.5443&-751.123&14.6187
  \end{bmatrix}.
\end{equation}
The chisquare for the reduced Planck data is defined as
\begin{equation}
\chi^2_{Planck}=X^TC_{Planck}^{-1}X,
\end{equation}
where
\begin{equation}
X=
  \begin{bmatrix}
  l_A-301.65\\
  R-1.7500\\
  z_*-1090.33.
  \end{bmatrix}
\end{equation}

\section{Results \label{sec:results}}

\begin{center}
 \begin{figure}
 \includegraphics[width=5in]{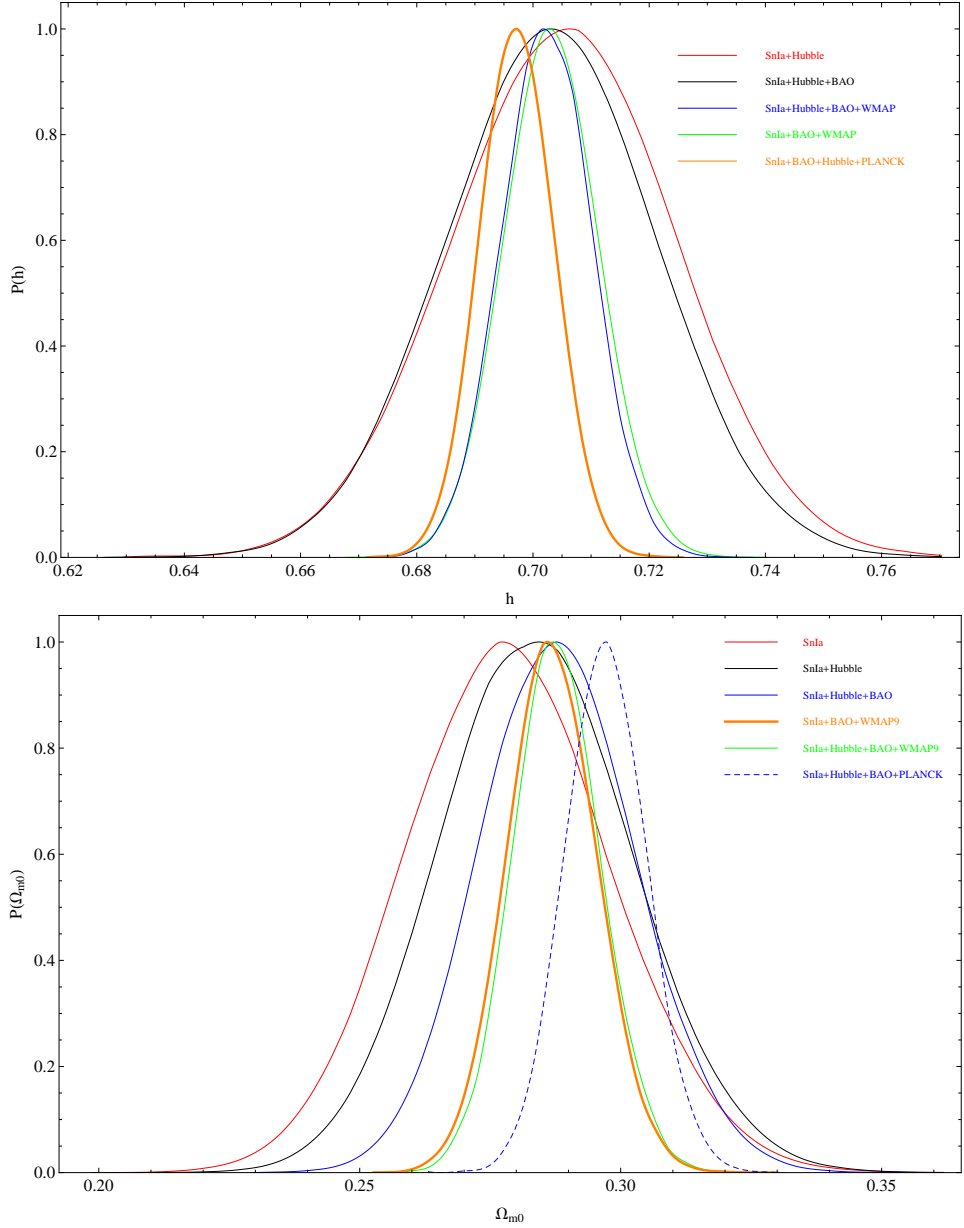}
 \caption{\label{fig:datasets comparing}Marginalized posterior distributions for $\Omega_{m0}$ (upper)
and $h$ (bottom) from SNe Ia data in combination with HUB, BAO
WMAP9 and Planck. }
\end{figure}
\end{center}

Using the Union$2.1$ sample in combination with other measurements
described in the previous section, we give the constraints on the
base $\Lambda$CDM model. The best-fit values of $\Omega_{m0}$ and
$h$ with $68\%$ CL errors are summarized in Table~\ref{tab2}, and
their likelihoods are shown in Figure~\ref{fig:datasets comparing}.

From Table~\ref{tab2} we can see that the SNe Ia data alone favor a
lower value of $\Omega_{m0}$ than the SNe Ia data in combination
with other datasets. Including the BAO and WMAP9/Planck data
improves significantly the constraint on $\Omega_{m0}$. Including
the reduced Planck data gives the highest $\Omega_{m0}$ and the lowest $h$.
We find these estimates of $\Omega_{m0}$ are consistent with each other
within 1$\sigma$ CL, but are in tension with the results derived by
Planck~\cite{Ade:2013zuv}. The estimates of $h$ from the HUB, BAO
and WMAP9 are compatible with those from Planck, but are discrepant
with those from fitting the calibrated SNe magnitude-redshift
relation~\cite{Riess:2011yx}.

\begin{table}[!th]
\begin{tabular}{l|l|l}
\hline ${\rm data}$  & $\Omega_{m0}$  & $h$  \tabularnewline \hline
\hline $SN$                       & $0.2776^{+0.0299}_{-0.0304}$ &
$-$ \tabularnewline \hline \hline $SN+HUB$                   &
$0.2834^{+0.0231}_{-0.0227}$ & $0.7055^{+0.0234}_{-0.0248}$
\tabularnewline \hline \hline $SN+HUB+BAO$               &
$0.2875^{+0.0192}_{-0.0192}$ & $0.7037^{+0.0241}_{-0.0248}$
\tabularnewline \hline \hline $SN+BAO+WMAP9$             &
$0.2867^{+0.0113}_{-0.0111}$ & $0.7032^{+0.0107}_{-0.0108}$
\tabularnewline \hline
\hline $SN+HUB+BAO+WMAP9$          & $0.2877^{+0.0111}_{-0.0110}$ &
$0.7022^{+0.0110}_{-0.0095}$ \tabularnewline \hline \hline
$SN+HUB+BAO+Planck$        & $0.2969^{+0.0103}_{-0.0086}$ &
$0.6974^{+0.0082}_{-0.0086}$ \tabularnewline \hline
\end{tabular}
\caption{\label{tab2}Constraints with 1$\sigma$ errors on
$\Omega_{m0}$ and $h$ for the base $\Lambda$CDM cosmology from SNe
Ia data in combination with HUB, BAO, WMAP9 and Planck.}
\end{table}

\begin{table}[!h]
\begin{tabular}{l|l|l}
\hline ${\rm redshift\, range}$  & $\Omega_{m0}$  & $h$
\tabularnewline \hline \hline $0-0.28$                 &
$0.3187^{+0.0787}_{-0.0932}$ & $0.6870^{+0.0381}_{-0.0298}$
\tabularnewline \hline \hline $0.28-0.73$              &
$0.3032^{+0.0332}_{-0.0327}$ & $0.6677^{+0.0556}_{-0.0469}$
\tabularnewline \hline \hline $>0.73(WMAP)$                  &
$0.2767^{+0.0366}_{-0.0328}$ & $0.7123^{+0.0319}_{-0.0345}$
\tabularnewline \hline \hline $>0.73(Planck)$                  &
$0.3158^{+0.0230}_{-0.0218}$ & $0.6831^{+0.0146}_{-0.0181}$
\tabularnewline \hline \hline ${\rm whole}(WMAP)$                &
$0.2877^{+0.0111}_{-0.0110}$ & $0.7022^{+0.0110}_{-0.0095}$
\tabularnewline \hline \hline ${\rm whole}(Planck)$        &
$0.2969^{+0.0103}_{-0.0086}$ & $0.6974^{+0.0082}_{-0.0086}$
\tabularnewline \hline
\end{tabular}
\caption{\label{tab:tomography}Constraints with 1$\sigma$ errors on
$\Omega_{m0}$ and $h$ for the base $\Lambda$CDM cosmology in three
redshift bins from the SN+HUB+BAO+WMAP9 and SN+HUB+BAO+Planck. For a
comparison, we also list the constraints in the whole redshift
range.}
\end{table}

\begin{center}
 \begin{figure}[!h]
 \centering
 \includegraphics[scale=0.85]{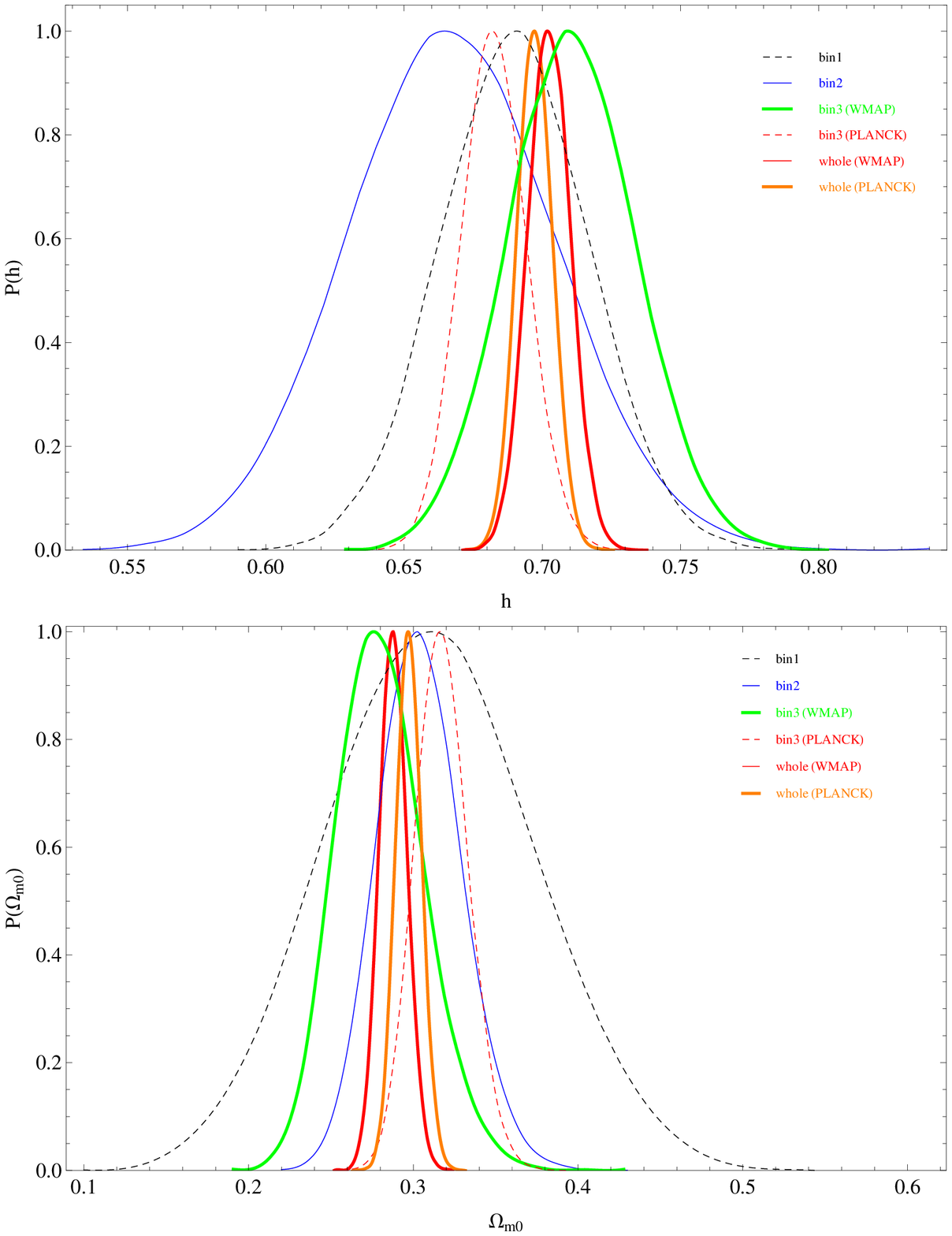}
 \caption{\label{fig:3bins}Marginalized posterior distributions of $\Omega_{m0}$ (upper)
and $h$ (bottom) for three redshift bins of the
SN+HUB+BAO+WMAP9/Planck data.}
\end{figure}
\end{center}

Using the SN+HUB+BAO+WMAP9/Planck data distributed in three
different redshift bins, we present the constraints on $\Omega_{m0}$
and $h$ for the $\Lambda$CDM model in Table~\ref{tab:tomography}. The
corresponding marginalized posterior distributions are shown in
Figure~\ref{fig:3bins}.

\begin{center}
\begin{figure}[!h]
  \includegraphics[width=5in]{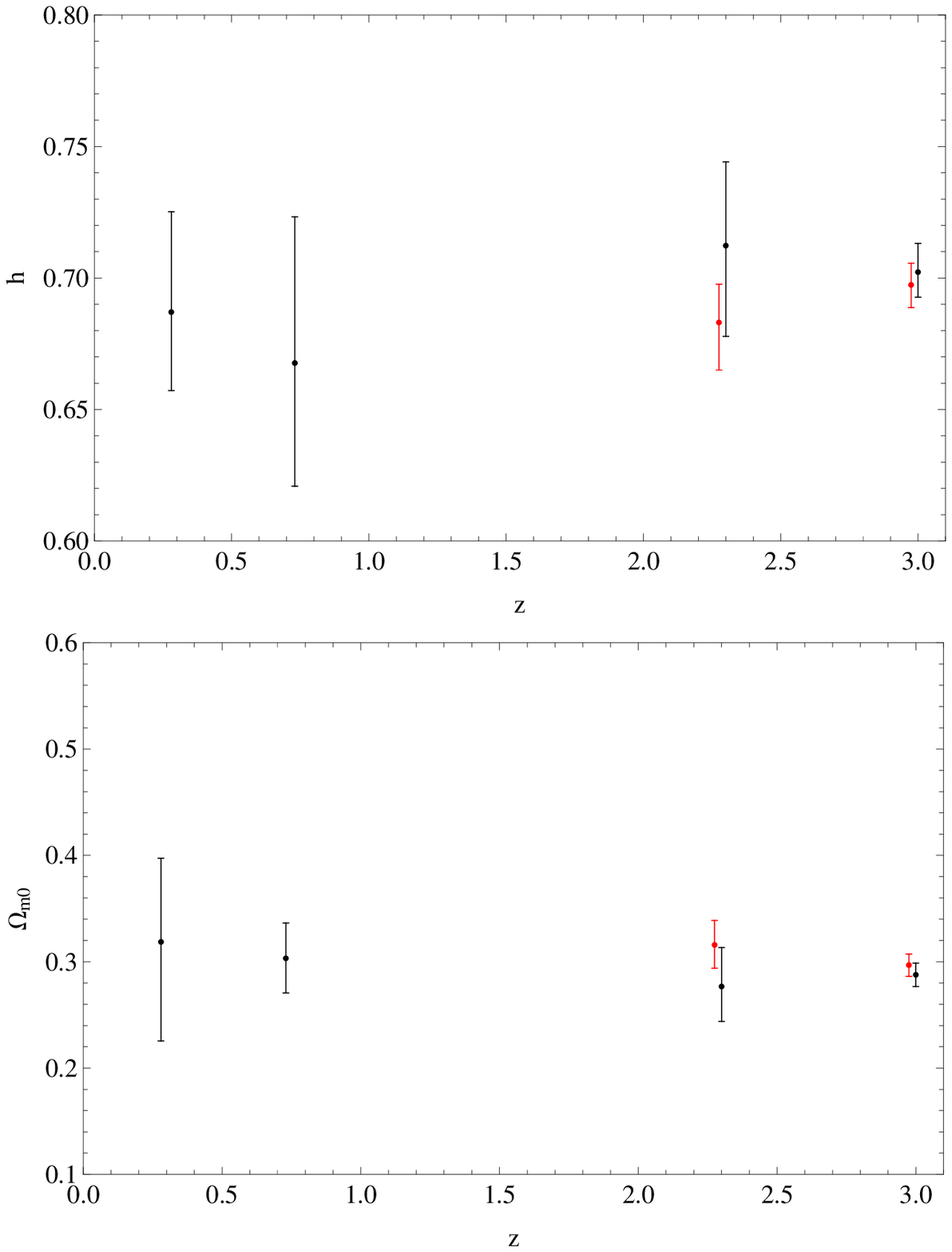}
  \caption{\label{fig:error} Best-fit values of $\Omega_{m0}$ (upper) and $h$ (bottom)
with $1\sigma$ errors in three redshift bins and in the whole
redshift range. The third and forth error bars in black represent
the case containing WMAP9 data, while the red ones represent the
case containing Planck data. }
\end{figure}
\end{center}

Our analysis shows that low-redshift observations give a higher
value of $\Omega_{m0}$,  while high-redshift
observations give a lower one by
using the SN+HUB+BAO+WMAP9 data. However, the high redshift
$z > 0.73$ observations with Planck data favor a relatively higher value of $\Omega_{m0}$,
which is inconsistent with the high-redshift value from WMAP9 at about
$1.1\sigma\ CL.$  In addition, there are large
uncertainties in the estimation of $\Omega_{m0}$ from the data in
the redshift range $0 < z < 0.28$. From Table~\ref{tab:tomography}
we find that the data in the mid-redshift range $0.28 < z < 0.73$
favor a lower Hubble constant with a little large uncertainty than the data at low and high
redshifts.  Figure~\ref{fig:error}
shows the best-fit values of $\Omega_{m0}$ and $h$ with 1$\sigma$
errors for the data in three different redshift bins.

In our analysis the Hubble constant is marginalized as a nuisance
parameter in the SNe Ia likelihood function. Therefore, the
constraints on $h$ mainly come from the HUB, BAO and WMAP9/Planck data. In
Ref.~\cite{Zhang:2013hma} a higher value of Hubble constant is
recently obtained from measurements of nearby SNe Ia with help of
measurements of Cepheid variables, than that obtained by Planck.
However, our estimates of $h$ from the data in the redshift ranges
of $z<0.28$ and $0.28<z<0.73$ are lower than the result obtained
in~\cite{Zhang:2013hma}. Moreover, the high-redshift data ($z>0.73$)
including the WMAP9 data favor a higher value of $h$ than the data
in the first two redshift bins.



%


\section{Conclusions \label{sec:conclusions}}

The estimates of $\Omega_{m0}$ and $h$ in the base $\Lambda$CDM model should be consistent with
each other from measurements made in different redshift intervals, if the simplest $\Lambda$CDM model
completely describes the evolution of our universe and the unknown sources of systematic errors of
these measurements can be negligible.
The recent Planck observations of the CMB lead to a Hubble constant of $h=0.673\pm0.012$ and
a matter density parameter of $\Omega_{m0}=0.315\pm0.017$~\cite{Ade:2013zuv}, which, however, are different from the low-$z$ measurements.
In this work, we have studied the consistency of the estimated values for the Hubble constant
and matter density parameter from different redshift data.

We have first obtained reduced CMB data for ${\{R, l_A, z_*\} }$ from WMAP9 and Planck data, based on a flat $\Lambda CDM $ model. We then have placed constraints on the base $\Lambda$CDM model using
astrophysical measurements of SNe Ia, Hubble and BAO, in combination
with the reduced WMAP9/Planck CMB data. We have found that the SNe Ia
data alone favor a lower value of $\Omega_{m0}$ and adding the HUB,
BAO, and the reduced WMAP9/reduced Planck data can give a higher one, but it is
still in tension with the result reported by Planck. Moreover, the
estimates of $h$ from the HUB, BAO and WMAP9 are compatible with
those from Planck, but are discrepant with those from fitting the
calibrated SNe magnitude-redshift relation~\cite{Riess:2011yx}. There is no any tension
on $h$ among three redshift bins, as shown in Fig.~\ref{fig:error}.

We have also implemented the redshift tomography analysis in the
context of the $\Lambda$CDM cosmology with the SNe Ia, HUB, BAO and
CMB data. We have found that low-redshift observations ($z < 0.28$)
give a higher value of $\Omega_{m0}$, as estimated by Planck, while
high-redshift observations ($z > 0.73$) with the WMAP9 data give a lower one, which is
inconsistent with that from the SN+HUB+BAO data in the high-redshift range in combination
with the Planck data at about $1.1\sigma\ CL$.  In
addition, the data in the mid-redshift range $0.28 < z < 0.73$ favor
a lower Hubble constant. The current data
cannot provide statistically significant evidence for any tension
among the different redshift bins.


\begin{acknowledgments}
This work was supported in part by the National Natural Science Foundation of China
(No.10821504, No.10975168, No.11035008 and No.11175225), and in part by the Ministry of Science and
Technology of China under Grant No. 2010CB833004 and No. 2010CB832805.
\end{acknowledgments}



\begin{thebibliography}{99}

\bibitem{Riess:1998cb}
  A.~G.~Riess {\it et al.}  [Supernova Search Team Collaboration],
  Astron.\ J.\  {\bf 116}, 1009 (1998)
  [astro-ph/9805201].

\bibitem{Perlmutter:1998np}
  S.~Perlmutter {\it et al.}  [Supernova Cosmology Project Collaboration],
  Astrophys.\ J.\  {\bf 517}, 565 (1999)
  [astro-ph/9812133].

\bibitem{Sahni:2002kh}
  V.~Sahni,
  Class.\ Quant.\ Grav.\  {\bf 19}, 3435 (2002)
  [astro-ph/0202076].

\bibitem{Padmanabhan:2002ji}
  T.~Padmanabhan,
  Phys.\ Rept.\  {\bf 380}, 235 (2003)
  [hep-th/0212290].

\bibitem{Suzuki:2011hu}
  N.~Suzuki, D.~Rubin, C.~Lidman, G.~Aldering, R.~Amanullah, K.~Barbary, L.~F.~Barrientos and J.~Botyanszki {\it et al.},
  Astrophys.\ J.\  {\bf 746}, 85 (2012)
  [arXiv:1105.3470 [astro-ph.CO]].

\bibitem{Hinshaw:2012aka}
  G.~Hinshaw {\it et al.}  [WMAP Collaboration],
  Astrophys.\ J.\ Suppl.\  {\bf 208}, 19 (2013)
  [arXiv:1212.5226 [astro-ph.CO]].


\bibitem{Ade:2013zuv}
  P.~A.~R.~Ade {\it et al.}  [Planck Collaboration],
  arXiv:1303.5076 [astro-ph.CO].

\bibitem{Riess:2011yx}
  A.~G.~Riess, L.~Macri, S.~Casertano, H.~Lampeitl, H.~C.~Ferguson, A.~V.~Filippenko, S.~W.~Jha and W.~Li {\it et al.},
  Astrophys.\ J.\  {\bf 730}, 119 (2011)
  [Erratum-ibid.\  {\bf 732}, 129 (2011)]
  [arXiv:1103.2976 [astro-ph.CO]].

\bibitem{Freedman:2012ny}
  W.~L.~Freedman, B.~F.~Madore, V.~Scowcroft, C.~Burns, A.~Monson, S.~E.~Persson, M.~Seibert and J.~Rigby,
  Astrophys.\ J.\  {\bf 758}, 24 (2012)
  [arXiv:1208.3281 [astro-ph.CO]].

\bibitem{Xia:2013dea}
  J.~-Q.~Xia, H.~Li and X.~Zhang,
  Phys.\ Rev.\ D {\bf 88}, 063501 (2013)
  [arXiv:1308.0188 [astro-ph.CO]].

\bibitem{Cheng:2013csa}
C.~Cheng and Q.~-G.~Huang,
  Phys.\ Rev.\ D {\bf 89}, 043003 (2014)  [arXiv:1306.4091 [astro-ph.CO]].  


 \bibitem{Hu:2013aqa}
  B.~Hu, M.~Liguori, N.~Bartolo and S.~Matarrese,
  Phys.\ Rev.\ D {\bf 88}, 123514 (2013)
  [arXiv:1307.5276 [astro-ph.CO]].

\bibitem{Li:2013dwy}
  Z.~Li, P.~Wu, H.~Yu and Z.~-H.~Zhu,
   Sci.\ China Phys.\ Mech.\ Astron.\  {\bf 57}, 381 (2014)  [arXiv:1311.3467 [astro-ph.CO]].  


\bibitem{Efstathiou:2013via}
  G.~Efstathiou,
  arXiv:1311.3461 [astro-ph.CO].

\bibitem{Union2.1}
   N.~Suzuki {\it et al.},
 Astrophys.\ J.\ {\bf 746}, 85 (2012)
  [arXiv:1105.3470 [astro-ph.CO]].


\bibitem{Zhang:2013hma}
  S.~-N.~Zhang and Y.~-Z.~Ma,
  arXiv:1303.6124 [astro-ph.CO].

\bibitem{Simon:2004tf}
  J.~Simon, L.~Verde and R.~Jimenez,
  Phys.\ Rev.\ D {\bf 71}, 123001 (2005)
  [astro-ph/0412269].

\bibitem{Stern:2009ep}
  D.~Stern, R.~Jimenez, L.~Verde, M.~Kamionkowski and S.~A.~Stanford,
  JCAP {\bf 1002}, 008 (2010)
  [arXiv:0907.3149 [astro-ph.CO]].

\bibitem{Moresco:2012jh}
  M.~Moresco, A.~Cimatti, R.~Jimenez, L.~Pozzetti, G.~Zamorani, M.~Bolzonella, J.~Dunlop and F.~Lamareille {\it et al.},
  JCAP {\bf 1208}, 006 (2012)
  [arXiv:1201.3609 [astro-ph.CO]].


  O.~Farooq and B.~Ratra,
  Astrophys.\ J.\  {\bf 766}, L7 (2013)
  [arXiv:1301.5243 [astro-ph.CO]].



\bibitem{Mangano:2005cc}
  G.~Mangano, G.~Miele, S.~Pastor, T.~Pinto, O.~Pisanti and P.~D.~Serpico,
  Nucl.\ Phys.\ B {\bf 729}, 221 (2005)
  [hep-ph/0506164].



\bibitem{Nesseris:2005ur}
  S.~Nesseris and L.~Perivolaropoulos,
  Phys.\ Rev.\ D {\bf 72}, 123519 (2005)
  [astro-ph/0511040].





\bibitem{Beutler:2011hx}
  F.~Beutler, C.~Blake, M.~Colless, D.~H.~Jones, L.~Staveley-Smith, L.~Campbell, Q.~Parker and W.~Saunders {\it et al.},
  Mon.\ Not.\ Roy.\ Astron.\ Soc.\  {\bf 416}, 3017 (2011)
  [arXiv:1106.3366 [astro-ph.CO]].

\bibitem{Eisenstein:1997ik}
  D.~J.~Eisenstein and W.~Hu,
  Astrophys.\ J.\  {\bf 496}, 605 (1998)
  [astro-ph/9709112].



\bibitem{Percival:2009xn}
  W.~J.~Percival {\it et al.}  [SDSS Collaboration],
  Mon.\ Not.\ Roy.\ Astron.\ Soc.\  {\bf 401}, 2148 (2010)
  [arXiv:0907.1660 [astro-ph.CO]].

\bibitem{Eisenstein:2006nk}
  D.~J.~Eisenstein, H.~-j.~Seo, E.~Sirko and D.~Spergel,
  Astrophys.\ J.\  {\bf 664}, 675 (2007)
  [astro-ph/0604362].

\bibitem{Padmanabhan:2012hf}
  N.~Padmanabhan, X.~Xu, D.~J.~Eisenstein, R.~Scalzo, A.~J.~Cuesta, K.~T.~Mehta and E.~Kazin,
  Mon.\ Not.\ Roy.\ Astron.\ Soc.\  {\bf 427}, no. 3, 2132 (2012)
  [arXiv:1202.0090 [astro-ph.CO]].


\bibitem{Anderson:2012sa}
  L.~Anderson, E.~Aubourg, S.~Bailey, D.~Bizyaev, M.~Blanton, A.~S.~Bolton, J.~Brinkmann and J.~R.~Brownstein {\it et al.},
  Mon.\ Not.\ Roy.\ Astron.\ Soc.\  {\bf 427}, no. 4, 3435 (2013)
  [arXiv:1203.6594 [astro-ph.CO]].

\bibitem{Blake:2011en}
  C.~Blake, E.~Kazin, F.~Beutler, T.~Davis, D.~Parkinson, S.~Brough, M.~Colless and C.~Contreras {\it et al.},
  Mon.\ Not.\ Roy.\ Astron.\ Soc.\  {\bf 418}, 1707 (2011)
  [arXiv:1108.2635 [astro-ph.CO]].

\bibitem{Gaztanaga:2008de}
  E.~Gaztanaga, R.~Miquel and E.~Sanchez,
  Phys.\ Rev.\ Lett.\  {\bf 103}, 091302 (2009)
  [arXiv:0808.1921 [astro-ph]].

\bibitem{Hu:1995en}
  W.~Hu and N.~Sugiyama,
  Astrophys.\ J.\  {\bf 471}, 542 (1996)
  [astro-ph/9510117].

\bibitem{Wang:2013mha}
  Y.~Wang and S.~Wang,
  Phys.\ Rev.\ D {\bf 88}, 043522 (2013)
  [arXiv:1304.4514 [astro-ph.CO]].


\bibitem{Shafer:2013pxa}
D.~L.~Shafer and D.~Huterer,
  Phys.\ Rev.\ D {\bf 89}, 063510 (2014)  [arXiv:1312.1688 [astro-ph.CO]].  




\end{thebibliography}
\end{document}